\documentstyle[prl,aps,epsfig,floats,twocolumn,tighten]{revtex}

\newcommand{\beq}{\begin{equation}}
\newcommand{\eeq}{\end{equation}}
\newcommand{\beqa}{\begin{eqnarray}}
\newcommand{\eeqa}{\end{eqnarray}}

\begin{document}
\draft
\twocolumn[\hsize\textwidth\columnwidth\hsize\csname @twocolumnfalse\endcsname
\title{Spectral Gamma-ray Signatures of Cosmological Dark Matter
Annihilations}
\author{Lars Bergstr\"om, Joakim Edsj\"o}
\address{Department of Physics, Stockholm University,
SCFAB, SE-106 91 Stockholm, Sweden}
\author{Piero Ullio}
\address{SISSA, via Beirut 4, 34014 Trieste, Italy}
\maketitle
 
\begin{abstract}
We propose a new signature for weakly interacting massive particle
(WIMP) dark matter, a spectral feature in the diffuse extragalactic
gamma-ray radiation. This feature, a sudden drop of the
gamma-ray intensity at an energy corresponding to the WIMP mass,
comes from the asymmetric distortion of the line due to WIMP
annihilation into two gamma-rays caused by the cosmological redshift.
Unlike other proposed searches for a line signal, this method is
not very sensitive to the exact dark matter density distribution
in halos and subhalos. 

\end{abstract}
 
]

It has been known since long that particle dark matter annihilations
may produce an observable gamma-ray line
\cite{STS,rudaz,BS,jungkam,BU,ub}.  One of the prime particle dark matter
candidates is a WIMP (Weakly Interacting Massive Particle), of which
the supersymmetric neutralino is a favourite example.  WIMP
annihilation into $\gamma \gamma$ and $Z \gamma$ would give
monochromatic gamma rays with an energy equal to (or close to) the
WIMP mass \cite{BS,BU,ub}. Since these gamma rays are monochromatic 
and have high 
energy they would constitute a spectacular signature of annihilating 
dark matter.

There has been a rapid development of the understanding of how 
structure forms in the Universe. In the current model for structure 
formation, the $\Lambda$CDM model, most of the matter is in the form 
of non-relativistic cold dark matter (CDM), but with a contribution 
to the energy density also from a cosmological constant 
($\Lambda$). As shown by detailed $N$-body simulations
(see, e.g., \cite{virgo,moore1} and references therein),
in such a picture large structures form by the successive merging of small
substructures, with smaller objects generally being denser.
The $N$-body simulations also show that the dark matter density profile
in clusters of galaxies and in single galaxies develops a steep cusp
near the center, $\rho_{CDM}(r)\sim r^{-\alpha}$ with $\alpha$ ranging
from $1$\cite{NFW} to $1.5$ \cite{moore2}.

At present, it is not clear whether these $N$-body predictions are in
agreement or not with available data. On large scales, this scenario
gives excellent agreement with observations, see, e.g., the prediction
of the Lyman-$\alpha$ absorption lines at high redshifts \cite{croft}.
On smaller scales, one of the main puzzles is how to properly include
baryonic matter. For instance, it appears that the contradiction
between the number of satellites found in the $N$-body simulation of a
halo with the size of the Milky Way and the number of those observed
may be explained by plausible mechanisms which make most small subhalos
dark~\cite{bullock}. It is less clear how to 
get agreement between the measured rotation curves of dwarf and low
surface brightness galaxies and those found
in $\Lambda$CDM simulations (see \cite{moorerev} for a recent review).

Here we will take the view that the $\Lambda$CDM picture is basically
correct and that structure forms hierarchically, with the number
density of halos of mass $M$ being distributed as $dN/dM\propto M^{-\beta}$
with $\beta\sim 1.9$ -- $2$, as predicted by Press-Schechter theory~\cite{ps}
and also verified in $N$-body simulations. Furthermore,
we will use that the concentration of halos grows with decreasing
mass.


Previous analyses
(e.g., \cite{silkstebb,BBU,begu,ted,beg,calcaneo}) have focused on the 
dark matter gamma-ray signals from the Galactic center and the 
halo of the Milky Way or isolated nearby galaxies and satellites; 
in these cases the actual dark matter distribution within halos plays 
a crucial role for the observability. The presence of substructures, 
as well as of central cusps, increases the detected rates
\cite{silkstebb,begu,ted,beg}, but still it may be difficult to resolve
such individual sources (see \cite{ted}). We now show
that the integrated signal of unresolved cosmological dark matter
halos gives a potential detection method which is more robust to
changes of the details of how the dark matter is distributed locally.


We consider
the lightest neutralino, $\chi$, of the MSSM (the Minimal 
Supersymmetric Standard Model) as our template particle. 
The mass range is from around 50 GeV up to several TeV (see
\cite{lberev} for a recent review).  
We start with the (unrealistic) case of all the dark matter being
smoothly distributed at all redshifts, and then modify the results by
introducing structure.  The comoving number density $n_c$ of
neutralinos, after decoupling from chemical equilibrium
(``freeze-out'') at very large temperatures ($T\sim m_\chi/20$) is
depleted slightly due to self-annihilations, governed by the Boltzmann
equation $ \dot n_c=-\langle\sigma v
\rangle(1+z)^3n_c^2,\label{eq:boltz}$ where $\langle\sigma v
\rangle$ is the thermally-averaged annihilation rate, which, to an
excellent approximation after freeze-out, is velocity independent,
since the neutralinos move non-relativistically.

Each of the $\chi$ particles that disappears
will give rise to $N_\gamma$ photons on the average, with an energy
distribution in the rest frame of the annihilation
\beq
{dN_\gamma(E)\over dE}={dN_{\rm cont}\over dE}(E)+b_{\gamma\gamma}
\delta\left(m_\chi-E\right),\label{eq:edist}
\eeq
where the first term gives the average continuum gamma ray distribution per
annihilating $\chi$ and the second term is the $\gamma\gamma$ line
contribution, with $b_{\gamma\gamma}$ being the branching ratio into
$\gamma\gamma$  (in this discussion, we neglect the $Z\gamma$
channel~\cite{ub}).
 
Gamma-rays observed today with an energy $E_0$ correspond
to an emitted energy $E=(1+z)E_0$. If we now track, using 
the Boltzman equation, the number of neutralinos
that have disappeared from redshift $z$
until now, and fold in the energy distribution Eq.~(\ref{eq:edist}),
we can estimate the diffuse extragalactic gamma ray flux.
Let $H_0$ be the Hubble parameter.
Using the relation between time and redshift (see, e.g.,
\cite{BGbook})
$d/dt=-H_0(1+z)h(z)d/dz$
with
$h(z)=\sqrt{\Omega_M(1+z)^3+\Omega_K(1+z)^2+\Omega_\Lambda}$,
where $\Omega_M$, $\Omega_\Lambda$ and $\Omega_K=
1-\Omega_M-\Omega_\Lambda$ are the present fractions of the critical density
given by matter, vacuum energy and curvature, the rate is
\beq
{dn_c(z)\over dz}=\kappa{(1+z)^2\over h(z)}n_c(z)^2\;,\label{eq:zboltz}
\eeq
where
$\kappa=\langle\sigma v\rangle/H_0$.
 
The differential spectrum of the number density $n_\gamma$ of photons
generated by annihilations is given by:
\beq
{dn_\gamma\over dz}=N_\gamma{dn_c\over dz}
=\int_0^{m_\chi}{dN_\gamma(E)\over dE}{dn_c\over dz}dE.
\eeq
Here, $dn_c/dz$ can be computed directly from (\ref{eq:zboltz}) to excellent
accuracy, replacing
the exact solution $n_c(z)$ by the present number density of neutralinos
$n_0$ on the right hand side. 
 

Approximating $\Omega_\chi \sim \Omega_M$, we obtain
$n_0=\rho_\chi/m_\chi=\rho_{\rm crit}\Omega_M/m_\chi$, where
$\rho_{\rm crit} =1.06\cdot 10^{-5}\ h^2\ {\rm GeV/cm}^3$ and $h$ is 
the Hubble parameter in units of 100 km s$^{-1}$ Mpc$^{-1}$, and
we find that the gamma-ray flux is given by:
\beqa
\phi_\gamma ={c\over 4\pi}{dn_\gamma\over dE_0}=8.3\cdot 10^{-14}
\ {\rm cm}^{-2}{\rm s}^{-1}{\rm sr}^{-1}{\rm GeV}^{-1}\times \nonumber\\
{\Gamma_{26}\Omega_M^2 h^3\over m^2_{100}}
\int_0^{z_{up}}dz {(1+z)^3e^{-z/z_{\max}}\over h(z)}
{dN_\gamma(E_0(1+z))\over dE}.\label{eq:master}
\eeqa
where we defined
$\Gamma_{26}=\langle\sigma v\rangle/(10^{-26}\ {\rm cm}^3{\rm s}^{-1})$ and
$m_{100}$ the mass in units of 100 GeV.
The term $e^{-z/z_{\max}}$ accounts for absorption of
gamma-rays along the line of sight. For the energies we are interested
in, $1$ GeV $\lesssim E_0\lesssim 500$ GeV, it is the starlight and
(poorly known) infrared background radiation which is the limiting factor.
An optical depth of order unity is reached for a redshift which can be 
approximated by $z_{\rm max}(E_0)\sim 3.3(E_0/10\ {\rm GeV})^{-0.8}$, 
which is a simplified representation of the results of \cite{salamon,primack}.
The exponential form is a good approximation for
small values of $z_{\rm max}$ as in most of our cases.
The upper limit of integration is $z_{up}=m_\chi/E_0-1$, since
the maximum rest frame energy of a photon in an annihilation event
is $E=m_\chi$. The gamma line contribution to (\ref{eq:master}) is
particularly simple, just picking out the integrand at $z+1=m_\chi/E_0$;
it has the very distinctive and potentially observable signature of being
asymmetrically smeared to lower energies (due to the redshift) and of
suddenly dropping just above $m_\chi$. The continuum emission
will produce a less conspicuous feature, a smooth ``bump'' below
one tenth of the neutralino mass which may be more difficult to detect.
To give an example of the results, we performed a large scan of the MSSM
parameter space obtained with the DarkSUSY package \cite{ds}. Models
with large $\gamma\gamma$ rates ($(\sigma v)_{2\gamma}\gtrsim 10^{-29}$
cm$^3$s$^{-1}$) exist in all the mass range from $m_\chi$ = 70 GeV
to several TeV. Consider a high-rate model with
$m_\chi=86$ GeV,  $\Gamma_{26}\sim 6$, $b_{\gamma\gamma}
\sim 3\cdot 10^{-3}$,
in the ``concordance'' cosmology $\Omega_M=0.3$, $\Omega_\Lambda=0.7$,
$h=0.65$ (see Ref.\,~\cite{beg} for the full range of predicted fluxes 
in the MSSM).
The continuous gamma-ray rest frame energy distribution per annihilating
particle comes mainly from hadronization and decay of $\pi^0$s and is
conveniently parametrized as 
${dN_{\rm cont}(E)/dE}=(0.42 / m_\chi)e^{-8x} / (x^{1.5}+0.00014)$
where $x=E/m_\chi$. The resulting flux near 86 GeV is around
$10^{-15}$ cm$^{-2}$s$^{-1}$sr$^{-1}$GeV$^{-1}$, some
five orders of magnitude below the diffuse extragalactic flux extrapolated
from the Energetic Gamma-Ray Experiment Telescope (EGRET) measurements
\cite{sreekumar}. 



We now include the important effects of structure formation.
Consider a halo of mass $M$ whose radial density profile can be
generically described by \cite{calcaneo}
$\rho_{DM}(r)=\rho'_{DM}f\left(r/a\right)$,
with $\rho'_{DM}$ being a characteristic density and $a$ a
length scale. These are found in $N$-body simulations
not to be independent parameters, as smaller halos are
associated with higher densities. 
We assume that the halo of mass $M$ accreted from a spherical
volume of radius $R_M$, determined by requiring that the average cosmological
density times that volume is equal to $M$, $\rho_0\cdot 4\pi R_M^3/3 =M$
(with $\rho_0\sim 1.3\cdot 10^{-6}\ {\rm GeV/cm}^{3}$).
The increase of average squared overdensity per halo
(which is what enters the annihilation rate) is given by:
\beqa
\Delta^2 &\equiv&
\langle\left({\rho_{DM}\over\rho_0}\right)^2\rangle_{r<R_M} = \left({\rho'_{DM}\over \rho_0}\right) {I_2\over I_1},
\eeqa
where $ I_n \equiv \int_0^{R_M/a}y^2dy(f(y))^n$.
Here the dependence on the upper limit of integration is rather weak,
while for the lower limit of $I_2$, in very singular profiles, like the
Moore profile,
a cut-off  has to be imposed due to rapid self-annihilations near
the center~\protect\cite{calcaneo}.
 
Computing $I_2/I_1$ numerically, and using values of 
$\rho'_{DM}/\rho_0$ as determined for Milky Way size halos from
\cite{calcaneo}, we find values of $\Delta^2$ of $2.3\cdot 10^{5}$
for the Moore profile, $1.5\cdot 10^{4}$ for the Navarro-Frenk-White (NFW)
profile\cite{NFW}, and $7\cdot 10^{3}$ for a cored, modified
isothermal profile (modified such that the density falls as $1/r^3$ at
large radii \cite{calcaneo}). 
The flux ratios, $30:2:1$ for these three models
should be compared with the ratios $1000:100:1$ obtained within
a 5-degree cone encompassing the galactic center \cite{calcaneo}.
 
We also take into account that the number density
of halos is scaling like $\sim 1/M^2$, and that small-mass halos
are denser. Again, we resort to the highest-resolution $N$-body
simulations available to date. Fitting the concentration parameter
of Moore-type halos by
\cite{calcaneo} $c\sim 100\, (M_{\rm vir}/h^{-1}M_\odot)^{-0.084}$, we
find to a good approximation
$\Delta^2\sim 2\cdot 10^{5}M_{12}^{-0.22}$,
where $M_{12}$ is the halo mass in units of $10^{12}$ solar masses.
This means that the total flux from a halo of mass $M$ scales
as $M^{0.78}$. Since the number density of halos goes as $M^{-2}$,
the fraction of flux coming from halos of mass $M$
scales as $M^{-1.22}$. Thus the gamma-ray flux will dominantly
come from the smallest CDM halos. In simulations, substructure has
been found on all scales (being limited only by numerical resolution).
For very small dark matter clumps, however, no gain in overdensity
is expected, since once the matter power spectrum enters
the $k^{-4}$ region a constant density is predicted \cite{ted}.
We conservatively set $10^5 - 10^6$ $M_\odot$ as the minimal scale.


Finally, regarding redshift dependences, we have assumed a constant
enhancement factor $\Delta^2$ out to $z\sim 1$, and have arbitrarily
imposed quadratic growth in the enhancement factor from $z\sim 10$ to
the fully developed epoch $z=1$.  (The observable feature is not
sensitive to this assumption.)  Furthermore, in Eq.~(\ref{eq:master})
we make the replacement $(1+z)^3 \rightarrow 1$ \cite{bullock550},
reflecting the fact that the we are now
dealing with a clumped, rather than a smooth distribution
with density scaling $\sim (1+z)^3$.

\begin{figure}[t]\begin{center}
\epsfig{file=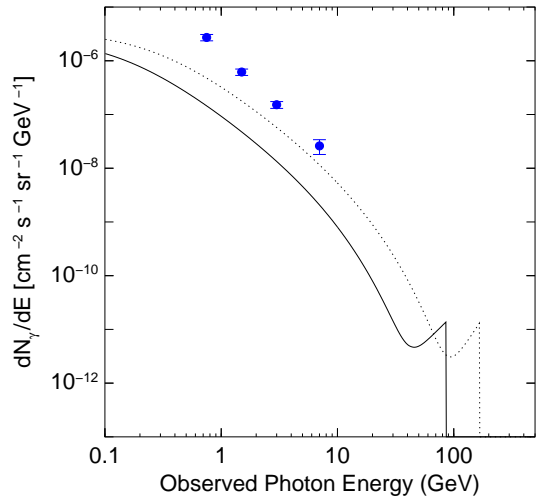,width=7cm}
\end{center}
\caption{The predicted diffuse $\gamma$-ray flux,
from cosmic annihilations into  continuum gamma-rays,
and a gamma-ray line. The redshifted line gives the 
conspicuous feature at the highest energies. Shown are cosmic
annihilation of 86 GeV  (solid line) and 166 GeV (dotted line)
neutralinos. A Moore density profile for the halo substructure
has been assumed. The EGRET data \protect\cite{sreekumar}
on the extragalactic flux are the data points with error bars shown.}
\label{fig:linewz}
\end{figure}

We thus arrive at the 
following expression for the flux including structure formation
\beqa
\phi_\gamma ={c\over 4\pi}{dn_\gamma\over dE_0}=8.3\cdot 10^{-14}
\ {\rm cm}^{-2}{\rm s}^{-1}{\rm sr}^{-1}{\rm GeV}^{-1}\times \nonumber\\
{\Gamma_{26}\Omega_M^2 h^3\over m^2_{100}}
\int_0^{z_{up}}dz {\Delta^2(z) e^{-z/z_{\max}}\over h(z)}
{dN_\gamma(E_0(1+z))\over dE}.\label{eq:master2}
\eeqa
We find that
the flux from small Moore-like halo structure is enhanced by roughly 
a factor $2\cdot 10^{6}$ compared to the smooth case, giving observability 
for the same annihilation parameters as used above. 
In Fig.\,~{\ref{fig:linewz}}, we show the results for the same 86 GeV 
MSSM model as in the example above, and a model of 166 GeV mass, 
$\Gamma_{26}=59$, $b_{\gamma\gamma}=1.2\cdot 10^{-4}$.

Several remarks are in order here.
 
(a) The flux calculated here is made up by the integrated effect of a large
    number of halos. One could also
detect nearby clumps in the Galactic halo \cite{calcaneo,beg}. This would
in particular make the line even more visible.
 
(b) Contrary to previous expectations (e.g., \cite{previous}), we do
not find it excluded that a large part of the measured extragalactic
gamma-ray flux around and above 10 GeV may originate from WIMP
annihilations.  Our more optimistic conclusion rely primarily on three
factors: (1) The improved understanding of CDM structure formation,
through semianalytic and numerical methods showing that less massive
halos are denser.  (2) The possibility that the density profile of
each subhalo can be steep, increasing further the emissivity.  (3) The
observability is helped by the existence of a sharp gamma-ray line in
the annihilation spectrum and the expected opacity of the universe to
this signal at the energies considered here, which gives a distinct
feature with no known astrophysical background.  None of these three
ingredients were present in the older analyses.

(c) An advantage of looking for a spectral feature
in the diffuse extragalactic gamma-ray background is that
regions of the sky can be used where foreground contamination
is as low as possible. Previous search strategies, focusing
on the Milky Way center or the center of other nearby galaxies,
are hampered by the relatively high gamma-ray flux
from other sources along these lines of sight.
 
(d)
We note that the
GLAST satellite \cite{glast} will cover the energy range up to 300 GeV with
unprecedented precision. It is possible that most of the measured
EGRET flux comes from Active Galactic Nuclei (AGN). Then, the absorption on
the infrared background will be severe above 80 GeV for all but the very
nearest AGNs. 
The true diffuse background above 80 GeV may therefore
be one or two orders of magnitude smaller than the EGRET extrapolation
$\sim E^{-2.1}$,
with corresponding higher chances of detecting a line signature.
Also note that the flux from the $2\gamma$ line and from
continuum gammas in neutralino annihilation are not strongly correlated
(except in the case of very pure higgsinos \cite{BBU}), so it may well
be that the line is visible but not the ``bump'', and vice versa.

(e) We notice that the spectral features of Fig.\,~{\ref{fig:linewz}
would appear with high significance, for the line it is 
of the order of 10$\sigma$
for a 5-year exposure with GLAST assuming the
background extrapolation $\sim E^{-2.1}$ given in \cite{sreekumar},
which becomes 20$\sigma$ if, 
in analogy with 
 the spectral slope of other cosmic rays,  
the background above 10 GeV drops like $E^{-2.7}$ instead.
Even if halos have less singular density profiles, such as
the cored isothermal sphere, a detectable signature may appear.
Alternatively, for
the Moore and NFW profiles, models with smaller gamma-ray
branching ratios and/or higher masses may be probed.
 
(f) The relatively broad spectral feature  caused by the redshifted
gamma-ray line alleviates the demand for very high energy resolution
of the detecting instrument put by the previous suggestions of
searching for the line in the local Galactic neighbourhood. We find
that an energy resolution of 10--20 \% is in fact adequate. For GLAST, this means
that the effective area can in fact be larger than we have assumed (since
a smaller requirement on the energy resolution means that events from a larger
part of the field of view of the detector can be used).
 
(g) It has to be reminded that the strength of the gamma-ray line
can be much lower than the examples we chose for Fig.\,~{\ref{fig:linewz},
in which case a detection would be correspondingly more difficult.
However, in many of those cases the continuum signal may be large.
In particular, we find that models compatible with the recent
results on the muon $g-2$ \cite{gm2} according to the suggested
contributions of ``new physics'' (such as supersymmetry) of \cite{marciano}
all give high values for the continuum gamma-ray flux.
 
(h) Although the clumpy structure is in the dark matter distribution,
and may not have optical counterparts, the rate of the annihilation
should scale with the overall matter distribution in the nearby Universe.
Thus, once mass maps from,
e.g., the Sloan Digital Sky Survey are available, a cross-correlation
analysis with the gamma-ray maps should enable a further reduction
of  Galactic gamma-ray foregrounds.
 
To conclude, we have suggested a new possible signature for
dark matter detection, which employs the clustering properties
of Cold Dark Matter as they emerge from $N$-body simulations.
The upcoming GLAST satellite, and possibly ground based
detectors, may have a good chance at detecting the characteristic
features, in particular the rise and sudden drop
in gamma-ray flux around the WIMP mass produced by annihilation
into two photons.
 
\section*{Acknowledgements}
L.B. and J.E. wish to thank the Swedish Research Council for 
financial support. P.U. was supported by the RTN project under 
grant HPRN-CT-2000-00152.
L.B. wants to thank the Institute for Advanced Study, Princeton,
for hospitality while this work was completed, and D. Spergel for
useful discussions.


\begin{references}
\bibitem{STS}
M.~Srednicki, S.~Theisen and J.~Silk,
Phys.\ Rev.\ Lett.\ {\bf 56}, 263 (1986); Erratum-ibid. {\bf 56},
1883 (1986).
\bibitem{rudaz}
S.~Rudaz,
Phys.\ Rev.\ Lett.\ {\bf 56}, 2128 (1986).
\bibitem{BS} L. Bergstr\"om and H. Snellman,
Phys.\ Rev.\ D {\bf 37}, 3737 (1988).
\bibitem{jungkam}
G.~Jungman and M.~Kamionkowski,
Phys.\ Rev.\ D {\bf 51}, 3121 (1995).
\bibitem{BU} L.~Bergstr\"om and P.~Ullio,
Nucl.\ Phys.\ B {\bf 504}, 27 (1997).
\bibitem{ub}
P.~Ullio and L.~Bergstr\"om, Phys.\ Rev.\ D {\bf 57}, 1962 (1998).
\bibitem{virgo}
A. Jenkins et al. (The Virgo Consortium),
Astrophys. J. {\bf 499}, 20 (1998).
\bibitem{moore1}
B.~Moore et al., Astrophys. J. {\bf 499}, L5 (1998).
\bibitem{NFW}J.F. Navarro, C.S. Frenk and S.D.M. White,
Astrophys. J. {\bf 462}, 563 (1996).
\bibitem{moore2}
S.~Ghigna et al., Astrophys. J. {\bf 544}, 616 (2000).
\bibitem{croft} R.A.C. Croft et al., Astrophys. J. {\bf 520}, 1 (1999).
\bibitem{bullock} J.S. Bullock, A.V. Kravtsov and D.H. Weinberg,
Astrophys. J. {\bf 548}, 33 (2001).
\bibitem{moorerev} B. Moore, 
astro-ph/0103100 (2001).
\bibitem{ps}
W.H.~Press and P.~Schechter, Astrophys. J. {\bf 187}, 425 (1974).
\bibitem{silkstebb} J. Silk and A. Stebbins, Astrophys. J. {\bf 411}, 439 (1993).
\bibitem{BBU} L. Bergstr\"om, J.H. Buckley and P. Ullio,
Astropart.\ Phys.\ {\bf 9}, 137 (1998).
\bibitem{begu} L. Bergstr\"om, J. Edsj\"o, P. Gondolo and P. Ullio,
Phys.\ Rev.\ D {\bf 59}, 043506 (1999).
\bibitem{ted}
E.~A.~Baltz, C.~Briot, P.~Salati, R.~Taillet and J.~Silk,
Phys.\ Rev.\ D {\bf 61}, 023514 (2000).
\bibitem{beg}
L. Bergstr\"om, J. Edsj\"o and C. Gunnarsson,
Phys.\ Rev.\ D {\bf 63}, 083515 (2001).
\bibitem{calcaneo} C. Calc\'aneo-Roldan and B. Moore,
Phys.\ Rev.\ D {\bf 62}, 123005 (2000).
\bibitem{lberev} L. Bergstr\"om,
Rept.\ Prog.\ Phys.\ {\bf 63}, 793 (2000).
\bibitem{BGbook} L. Bergstr\"om and A. Goobar, {\em Cosmology
and Particle Astrophysics}, Wiley/Praxis (Chichester), 1999.
\bibitem{salamon} M.H. Salamon and F.W. Stecker, Astrophys. J. {\bf 493}, 547 (1998).
\bibitem{primack} J.R. Primack, R.S. Somerville, J.S. Bullock and
J.E.G. Devriendt, eprint astro-ph/0011475 (2000).
\bibitem{ds} DarkSUSY package,
homepage\\ http://www.physto.se/\~{ }edsjo/darksusy. See
P.~Gondolo, J.~Edsj\"o, L.~Bergstr\"om, P.~Ullio and E.~A.~Baltz,
eprint astro-ph/0012234 (2000).

\bibitem{sreekumar} P. Sreekumar et al.,
Astrophys. J. {\bf 494}, 523 (1998).
\bibitem{bullock550}
J.S.~Bullock et al., Astrophys.\ J.\ {\bfseries 550}, 21 (2001). 
\bibitem{previous} J.E. Gunn, B.W. Lee, I. Lerche, D.N. Schramm and
G. Steigman, Astrophys. J. {\bf 223}, 1015 (1978);
F.W. Stecker, Astrophys. J. {\bf 223}, 1032 (1978);
J.S. Silk and M. Srednicki, Phys. Rev. Lett. {\bf 53}, 624 (1984);
Y.-T. Gao, F.W. Stecker and D.B. Cline, Astronomy and Astrophys. {\bf 249},
1 (1991).
\bibitem{glast} Gamma-ray Large Area Space Telescope (GLAST),
homepage http://www-glast.stanford.edu.
\bibitem{gm2} H.N. Brown et al., Phys. Rev. Lett. {\bf 86}, 2227 (2001).
\bibitem{marciano} W.J. Marciano and A. Czarnecki,
eprint hep-ph/0102122 (2001).

\end{references}
\end{document}